\documentclass[prd,aps,floatfix,nofootinbib,11 pt]{revtex4}
\usepackage{amssymb}
\usepackage{amsmath,graphicx,color,epsfig}

\input{tcilatex}

\begin{document}

\title{Decoherence and multipartite entanglement of non-inertial observers}
\author{M. Ramzan \thanks{%
mramzan@phys.qau.edu.pk}}

\address{Department of Physics Quaid-i-Azam University \\
Islamabad 45320, Pakistan}

\date{\today }

\begin{abstract}
Decoherence effect on multipartite entanglement in non-inertial frames is
investigated. GHZ state is considered to be shared between the partners with
one partner\ in inertial frame whereas the other two in accelerated frames.
One-tangle and $\pi $-tangles are used to quantify the entanglement of the
multipartite system influenced by phase damping and phase flip channels. It
is seen that for phase damping channel, entanglement sudden death (ESD)
occurs for $p>0.5$ in the infinite acceleration limit. On the other hand, in
case of phase flip channel, ESD behaviour happens around $50\%$ decoherence.
It is also seen that entanglement sudden birth (ESB) does occur in case of
phase flip channel. Furthermore, it is seen that effect of environment on
multipartite entanglement is much stronger than that of the acceleration of
non-inertial frames.\newline
\end{abstract}

\maketitle

\address{Department of Physics Quaid-i-Azam University \\
Islamabad 45320, Pakistan}


\vspace*{1.0cm}

\vspace*{1.0cm}

%\newpage

\section{Introduction}

Quantum entanglement has been recognized as a powerful tool for manipulating
information. The emerging field of quantum information processing has opened
a new way of using entanglement for performing tasks that are impossible to
achieve as efficiently with classical technologies. Quantum information and
quantum computation can process multiple tasks which are intractable with
classical technologies. Quantum entanglement is no doubt a fundamental
resource for a variety of quantum information processing tasks, such as
super-dense coding, quantum cryptography and quantum error correction [1-4].
During recent years, two important features entanglement sudden death (ESD)
and entanglement Sudden Birth (ESB) has been investigated [5-8]. Yu and
Eberly have investigated loss of entanglement in a finite time under the
action of pure vacuum noise for a bipartite system [9,10]. They found that,
even though it takes infinite time to complete decoherence locally, the
global entanglement may be lost in finite time. Multipartite entangled
states can be used to construct variety of information transmission
protocols, for example, quantum key distribution [11-12]. For the purposes
of quantum communication multipartite entangled states can serve as quantum
communication channels in most of the known teleportation protocols [13].
Furthermore, multipartite entanglement displays one of the most fascinating
features of quantum theory that is called as nonlocality of the quantum
world [14].

However, entangled states are very fragile when they are exposed to
environment. Decoherence is the major enemy of entanglement which is
responsible for emergence of classical behaviour in quantum systems [15].
Therefore, it could be of great importance to study the deteriorating effect
of decoherence in entangled states. Recently, researchers have focused on
relativistic quantum information in the field of quantum information science
due to conceptual and experimental reasons. In the last few years, much
attention had been given to the study of entanglement shared between
inertial and non-inertial observers by discussing how the Unruh or Hawking
effect will influence the degree of entanglement [16--27]. Implementation of
decoherence in non-inertial frames have been investigated for a two-qubit
system by Wang et. al [28]. Recently, multipartite entanglement in
non-inertial frames has been investigated [29-32], where it is shown that
entanglement is degraded by the acceleration of the inertial observers. Its
extension to the case of a qubit-qutrit system in non-inertial frames under
decoherence can be seen in ref. [33]. It is shown that ESB occurs in case of
depolarizing channel.

In this letter, the effect of decoherence is investigated for a multipartite
system in non-inertial frames by using phase damping and phase flip
channels. Three observers Alice, Bob and Charlie share a GHZ type state in
non-inertial frames. Alice is considered to be stationary whereas the other
two observers move with a uniform acceleration. One-tangles and $\pi $%
-tangles are calculated and discussed for both the channels. Two important
features of entanglement, ESD and ESB are investigated.

Let the three observers: Alice, an inertial observer, Bob and Charlie, the
accelerated observers moving with uniform acceleration, share the following
maximally entangled GHZ state%
\begin{equation}
\left\vert \Psi \right\rangle _{ABC}=\frac{1}{\sqrt{2}}\left. \left(
|0_{\omega _{a}}\rangle _{A}|0_{\omega _{b}}\rangle _{B}|0_{\omega
_{c}}\rangle _{C}+|1_{\omega _{a}}\rangle _{A}|1_{\omega _{b}}\rangle
_{B}|1_{\omega _{c}}\rangle _{C}\right) \right.
\end{equation}%
where $|0_{\omega _{a(bc)}}\rangle _{A(BC)}$ and $|1_{\omega
_{a(bc)}}\rangle _{A(BC)}$ are vacuum states and the first excited states
from the perspective of an inertial observer. Let the Dirac fields, as shown
in Refs. [27, 34-35], from the perspective of the uniformly accelerated
observers, are described as an entangled state of two modes monochromatic
with frequency $\omega _{i},$ $\forall _{i}$%
\begin{equation}
|0_{\omega _{i}}\rangle _{M}=\cos r_{i}|0_{\omega _{i}}\rangle
_{I}|0_{\omega _{i}}\rangle _{II}+\sin r_{i}|1_{\omega _{i}}\rangle
_{I}|1_{\omega _{i}}\rangle _{II}
\end{equation}%
and the only excited state is%
\begin{equation}
|1_{\omega _{i}}\rangle _{M}=|1_{\omega _{i}}\rangle _{I}|0_{\omega
_{i}}\rangle _{II}
\end{equation}%
where $\cos r_{i}=(e^{-2\pi \omega c/a_{i}}+1)^{-1/2}$, $a_{i}$ is the
acceleration of $i^{\text{th}}$ observer. The subscripts $I$ and $II$ of the
kets represent the Rindler modes in region $I$ and $II$, respectively, in
the Rindler spacetime diagram (see Ref. [28], Fig. (1)). Considering that an
accelerated observer in Rindler region $I$ has no access to the field modes
in the causally disconnected region $II$ and by taking the trace over the
inaccessible modes, one obtains the following tripartite state%
\begin{eqnarray}
\left\vert \Psi \right\rangle _{ABC} &=&\frac{1}{\sqrt{2}}[\cos r_{b}\cos
r_{c}|0\rangle _{A}|0\rangle _{B}|0\rangle _{C}+\cos r_{b}\sin
r_{c}|0\rangle _{A}|0\rangle _{B}|1\rangle _{C}  \notag \\
&&+\sin r_{b}\cos r_{c}|0\rangle _{A}|1\rangle _{B}|0\rangle _{C}+\sin
r_{b}\sin r_{c}|0\rangle _{A}|1\rangle _{B}|1\rangle _{C}+|0\rangle
_{A}|1\rangle _{B}|1\rangle _{C}]  \notag \\
&&
\end{eqnarray}%
For the sake of simplicity, the frequency subscripts are dropped and in
density matrix formalism, the above state can be written as%
\begin{eqnarray}
\rho _{AB_{I}CI} &=&\frac{1}{\sqrt{2}}[\cos r_{b}^{2}\cos
r_{c}^{2}|000\rangle \left\langle 000\right\vert +\cos r_{b}^{2}\sin
r_{c}^{2}|001\rangle \left\langle 001\right\vert   \notag \\
&&+\sin r_{b}^{2}\cos r_{c}^{2}|010\rangle \left\langle 010\right\vert +\sin
r_{b}^{2}\sin r_{c}^{2}|011\rangle \left\langle 011\right\vert   \notag \\
&&+\cos r_{b}\cos r_{c}(|000\rangle \left\langle 111\right\vert +|111\rangle
\left\langle 000\right\vert )+|111\rangle \left\langle 111\right\vert ]
\end{eqnarray}%
In order to simplify the calculations, it is assumed that Bob and Charlie
move with the same acceleration, i.e. $r_{b}=r_{c}=r.$ The well known
entanglement measure for a bipartite system, the negativity can be defined
as [36]%
\begin{equation}
\mathcal{N}_{AB}=\left\Vert \rho _{AB}^{T_{\alpha }}\right\Vert -1
\end{equation}%
where $T_{\alpha }$ is the partial transpose of $\rho _{AB}$ and $\left\Vert
.\right\Vert $ is the trace norm of a matrix. Whereas for a 3-qubit system $%
\left\vert \Psi \right\rangle _{ABC}$, $\mathcal{N}_{AB}$ defines the
two-tangle which is the negativity of the mixed state $\rho
_{AB}=Tr_{C}(\left\vert \Psi \right\rangle _{ABC}\left\langle \Psi
\right\vert )$ and its one-tangle can be defined by%
\begin{equation}
\mathcal{N}_{A(BC)}=\left\Vert \rho _{ABC}^{T_{\alpha }}\right\Vert -1
\end{equation}%
and the $\pi $-tangle is given by%
\begin{equation}
\mathcal{\pi }_{ABC}=\frac{1}{3}(\mathcal{\pi }_{A}+\mathcal{\pi }_{B}+%
\mathcal{\pi }_{C})
\end{equation}%
where $\mathcal{\pi }_{A(BC)}$ is the residual entanglement and is defined as%
\begin{equation}
\mathcal{\pi }_{A}=\mathcal{N}_{A(BC)}^{2}-\mathcal{N}_{AB}^{2}-\mathcal{N}%
_{AC}^{2}
\end{equation}

The interaction between the system and its environment introduces the
decoherence to the system, which is a process of the undesired correlation
between the system and the environment. The evolution of a state of a
quantum system in a noisy environment can be described by the super-operator
$\Phi $ in the Kraus operator representation as [37]

\begin{equation}
\rho _{f}=\Phi (\rho _{i})=\sum_{k}E_{k}\rho _{i}E_{k}^{\dag }  \label{E5}
\end{equation}%
where the Kraus operators $E_{i}$ satisfy the following completeness relation

\begin{equation}
\sum_{k}E_{k}^{\dag }E_{k}=I  \label{5}
\end{equation}%
We have constructed the Kraus operators for the evolution of the tripartite
system from the single qubit Kraus operators by taking their tensor product
over all $n^{3}$ combinations of $\pi \left( i\right) $ indices

\begin{equation}
E_{k}=\underset{\pi }{\otimes }e_{\pi \left( i\right) }  \label{6}
\end{equation}%
where $n$ correspond to the number of Kraus operators for a single qubit
channel. The single qubit Kraus operators for phase damping channel are
given by
\begin{equation}
E_{0}=\left[
\begin{array}{cc}
1 & 0 \\
0 & \sqrt{1-p}%
\end{array}%
\right] ,\quad E_{1}=\left[
\begin{array}{cc}
1 & 0 \\
0 & \sqrt{p}%
\end{array}%
\right]
\end{equation}%
and for phase flip channel
\begin{equation}
E_{0}=\sqrt{1-p}I,\quad E_{1}=\sqrt{p}\sigma _{z}
\end{equation}%
where $\sigma _{z}$ represents the usual Pauli matrix. Using equations
(7-14) along with the initial density matrix of as given in equation (5),
the one-tangles and $\pi $-tangles of the tripartite system under different
environments can be found as given in the following subsections.

The three one-tangles, influenced by the phase damping channel can be
calculated by using the definition as given in equation (7), are given by
\begin{eqnarray}
\mathcal{N}_{A(BC)} &=&-\frac{1}{2}+\frac{1}{2}\cos ^{4}r+\frac{1}{2}%
((1-p_{0})(1-p_{1})(1-p_{2})\cos ^{4}r)^{1/2}  \notag \\
&&+\frac{1}{2}((1-p_{0})(1-p_{1})(1-p_{2})\cos ^{4}r+\sin ^{8}r)^{1/2}+\frac{%
1}{4}\sin (2r)^{2}  \notag \\
&&
\end{eqnarray}%
and%
\begin{eqnarray}
\mathcal{N}_{B(AC)} &=&\mathcal{N}_{C(AB)}=  \notag \\
&&-\frac{1}{16}+\frac{1}{2}((1-p_{0})(1-p_{1})(1-p_{2})\cos ^{4}r)^{1/2}+%
\frac{1}{16}\cos (4r)  \notag \\
&&+\frac{1}{8}((16-16p_{0})(1-p_{1})(1-p_{2})\cos ^{4}r+\sin ^{4}(2r))^{1/2}
\end{eqnarray}%
where $p_{0}$, $p_{1}$ and $p_{2}$ are the decoherence parameters
corresponding to local coupling of the channel with the qubits of Alice, Bob
and Charlie, respectively. The collective coupling corresponds to the
situation when $p_{0}=p_{1}=p_{2}=p.$ The $\pi $-tangle can be calculated
easily by using equation (8) and is given by%
\begin{eqnarray}
\mathcal{\pi }_{ABC} &=&\frac{1}{3}(-\frac{1}{2}+\frac{1}{2}\cos ^{4}r+\frac{%
1}{2}((1-p_{0})(1-p_{1})(1-p_{2})\cos ^{4}r)^{1/2}  \notag \\
&&+\frac{1}{2}((1-p_{0})(1-p_{1})(1-p_{2})\cos ^{4}r+\sin ^{8}r)^{1/2}+\frac{%
1}{4}\sin (2r)^{2})^{2}  \notag \\
&&+\frac{2}{3}(-\frac{1}{16}+\frac{1}{2}((1-p_{0})(1-p_{1})(1-p_{2})\cos
^{4}r)^{1/2}+\frac{1}{16}\cos (4r)  \notag \\
&&+\frac{1}{8}((16-16p_{0})(1-p_{1})(1-p_{2})\cos ^{4}r+\sin
(2r)^{4})^{1/2})^{2}
\end{eqnarray}

The eigenvalues of the partial transpose matrix when only qutrit is
influenced by the depolarizing channel are given by%
\begin{eqnarray}
\mathcal{N}_{A(BC)} &=&-\frac{1}{2}+\frac{1}{2}\cos ^{2}r(\text{abs}%
[(1-2p_{0})(1-2p_{1})(1-2p_{2})]+\cos ^{2}r)  \notag \\
&&+\frac{1}{2}((1-2p_{0})^{2}(1-2p_{1})^{2}(1-2p_{2})^{2}\cos ^{4}r+\sin
^{8}r)^{1/2}+\frac{1}{4}\sin ^{2}(2r)  \notag \\
&&
\end{eqnarray}

\begin{eqnarray}
\mathcal{N}_{B(AC)} &=&\mathcal{N}_{C(AB)}=  \notag \\
&&-\frac{1}{2}+\frac{1}{2}\text{abs}[(1-2p_{0})(1-2p_{1})(1-2p_{2})]\cos
^{2}r+1/2\cos ^{4}r+\frac{1}{2}\sin ^{4}r  \notag \\
&&+\frac{1}{8}\sin (2r)^{2}+\frac{1}{8}%
(16(1-2p_{0})^{2}(1-2p_{1})^{2}(1-2p_{2})^{2}\cos ^{4}r+\sin ^{4}(2r))^{1/2}
\notag \\
&&
\end{eqnarray}%
The $\pi $-tangle can be calculated easily by using equation (8) and is
given by%
\begin{eqnarray}
\mathcal{\pi }_{ABC} &=&\frac{1}{3}(-\frac{1}{2}+\frac{1}{2}\cos ^{2}r(\text{%
abs}[(1-2p_{0})(1-2p_{1})(1-2p_{2})]+\cos ^{2}r)  \notag \\
&&+\frac{1}{2}((1-2p_{0})^{2}(1-2p_{1})^{2}(1-2p_{2})^{2}\cos ^{4}r+\sin
^{8}r)^{1/2}+\frac{1}{4}\sin (2r)^{2})^{2}  \notag \\
&&+\frac{2}{3}(-\frac{1}{2}+\frac{1}{2}\text{abs}%
[(1-2p_{0})(1-2p_{1})(1-2p_{2})]\cos ^{2}r+\frac{1}{2}\cos ^{4}r+\frac{1}{2}%
\sin ^{4}r  \notag \\
&&+\frac{1}{8}\sin (2r)^{2}+\frac{1}{8}%
(16(1-2p_{0})^{2}(1-2p_{1})^{2}(1-2p_{2})^{2}\cos ^{4}r+\sin
^{4}(2r))^{1/2})^{2}  \notag \\
&&
\end{eqnarray}%
The two-tangles $\mathcal{N}_{AB},$ $\mathcal{N}_{BC}$\ and $\mathcal{N}%
_{AC} $ can be easily calculated by taking the partial trace of the final
density matrix after the environmental effects over qubits $C,$ $A$ and $B$
respectively. All the two-tangles remain zero as expected since the reduced
density matrix is not affected by the environment.

Analytical expressions for one-tangles and $\pi $-tangles are calculated for
a multipartite system in non-inertial frames influenced by phase damping and
phase flip environments. The results are consistent with refs. [30, 31] and
can be easily verified from the expressions of one-tangles and $\pi $%
-tangles. It is seen that for $r=\pi /4$ all the one-tangles become equal,
i.e. $\mathcal{N}_{A(BC)}=\mathcal{N}_{B(AC)}=\mathcal{N}_{C(AB)}$ for both
the environments under consideration$.$ To investigate the effect of
decoherence on the multipartite system, the one-tangles and $\pi $-tangles
are plotted as a function of decoherence parameter, $p$ for different values
of acceleration $r$\ for phase damping channel in figure 1. Figure 1 (a-c)
show the behavior of one-tangles and $\pi $-tangles when only Alice's qubit
is coupled to the phase damping channel. Whereas figures 1 (d-e) show the
behavior of one-tangles and $\pi $-tangles when all the three qubits are
coupled to the phase damping channel. It is seen that the $\pi $-tangles are
heavily influenced by the environment as compared to the one-tangles for
both the cases (local and collective couplings). Furthermore, as the value
of acceleration $r,$ entanglement degradation is enhanced and it is more
prominent for higher values of decoherence parameter $p$. Since, a similar
behavior of one-tangles ($\mathcal{N}_{A(BC)},$ and $\mathcal{N}_{B(AC)}$)
is seen therefore, only $\mathcal{N}_{A(BC)}$ is plotted for discussion. It
is shown that the one-tangles and $\pi $-tangles goes to zero at $p=1$ for
phase damping channel. On the other hand, it goes to zero at $p=0.5$ in case
of phase flip channel.

In figure 2, the one-tangles and $\pi $-tangles are plotted as a function of
decoherence parameter, $p$ for different values of acceleration $r$\
influenced by the phase flip channel. Similar to the figure 1, figure 2
(a-c) show the behavior of one-tangles and $\pi $-tangles when only Alice's
qubit is coupled to the phase flip channel. Whereas figures 2 (d-e) show the
behavior of one-tangles and $\pi $-tangles when all the three qubits are
coupled to the phase flip channel. It is seen that the $\pi $-tangles are
heavily influenced by the environment in both the cases (local and
collective couplings). It is seen that maximum entanglement degradation
occurs at $p=0.5$ irrespective of the value of acceleration $r$. It is seen
that a sudden entanglement rebound process takes place for $p>0.5$ in the
case of phase flip noise. It is also seen that the one-tangles and $\pi $%
-tangles go to zero at $p=1$. Since the rebound process is much more
stronger than the resistance of acceleration, one cannot ignore it
anyway as it is much more prominent for $p>0.75$. In Fig. 3, the
three-dimensional graphs for one-tangles and $\pi $-tangles are
given as a function of decoherence parameter, $p$ and acceleration
$r$ for phase damping and phase flip channels. It is shown that
different environments affect the entanglement of the tripartite
system differently. It is also seen that in the case of the phase
flip channel, ESD behavior becomes independent of the acceleration.
However, the sudden death of entanglement cannot be avoided for a
phase flip noise around 50\% decoherence. Furthermore, entanglement
dies out more quickly compared to the phase damping channel for a
lower level of decoherence in the case of phase flip noise. The
sudden birth and non-vanishing behavior of the one-tangles $\pi
$-tangle at infinite acceleration is an interesting result. Since
Rindler spacetime is similar to Schwarzschild spacetime, it enables
one to conjecture that multipartite entanglement does not vanish
even if one party falls into the event horizon of the black hole.
Hence, some quantum information processing, for example,
teleportation, [38] can be performed within and outside the black
hole.

In summary, environmental effects on tripartite entanglement in non-inertial
frames is investigated by considering a maximally entangled GHZ state shared
between three partners. It is assumed that the two partners are in
accelerated frames moving with same uniform acceleration. In order to
investigate the environmental effects on the entanglement, one-tangles and $%
\pi $-tangles are calculated for the multipartite system. It is seen that in
case of phase damping channel, entanglement sudden death (ESD) occurs for
higher values of decoherence. Whereas, for the phase flip channel, ESD
behaviour happens at $p=0.5$. In addition, prominent behaviour of
entanglement sudden birth (ESB) is seen for $p>0.75$ in case of phase flip
channel. Therefore, it is investigated that the effect of environment is
much stronger than that of acceleration for multipartite systems.

\begin{figure}[tbp]
\begin{center}
\vspace{-2cm} \includegraphics[scale=0.8]{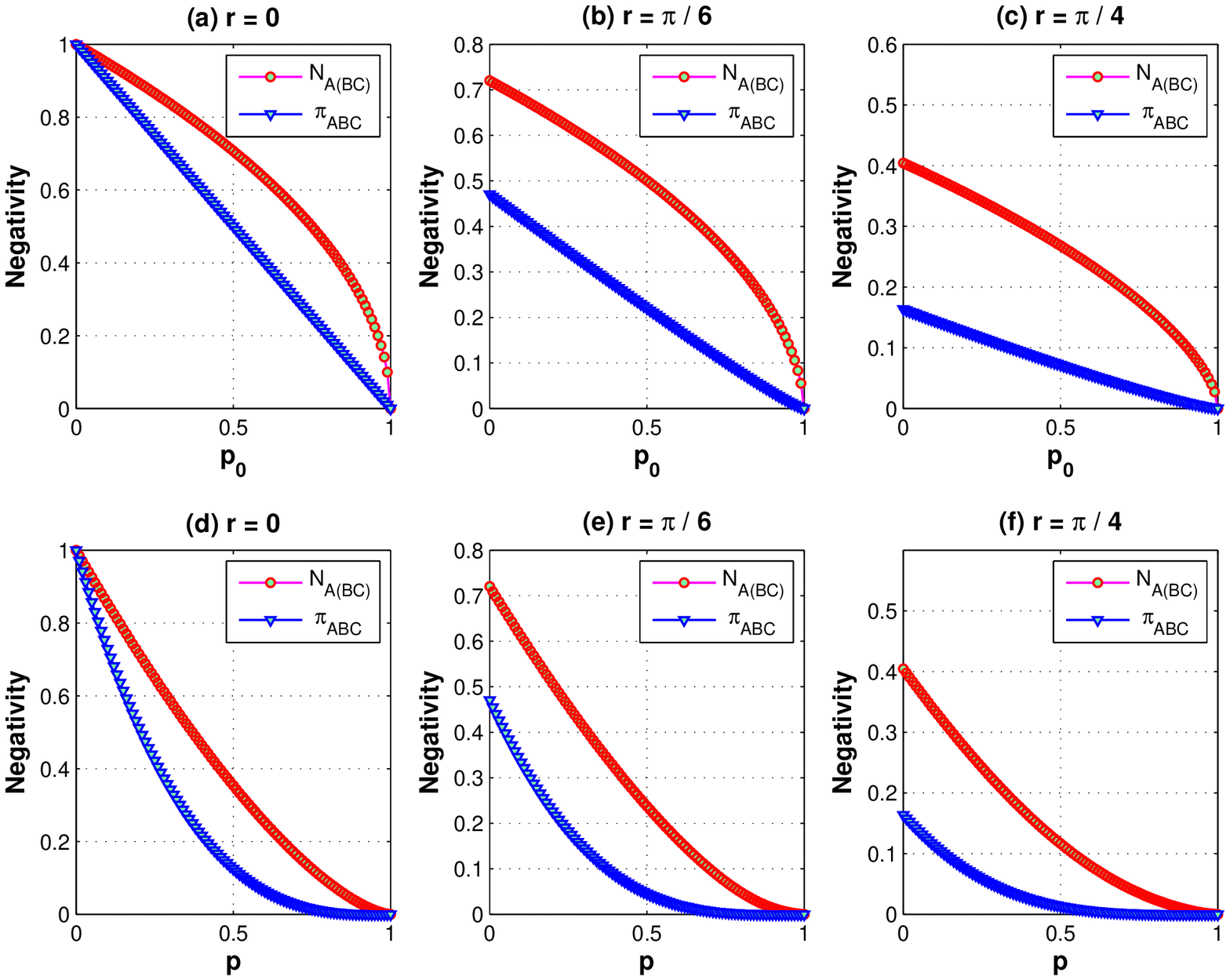} \\[0pt]
\end{center}
\caption{(Color online). The one-tangles and $\protect\pi $-tangles are
plotted as a function of decoherence parameter, $p$ for different values of
acceleration $r$\ for phase damping channel.}
\end{figure}

\begin{figure}[tbp]
\begin{center}
\vspace{-2cm} \includegraphics[scale=0.8]{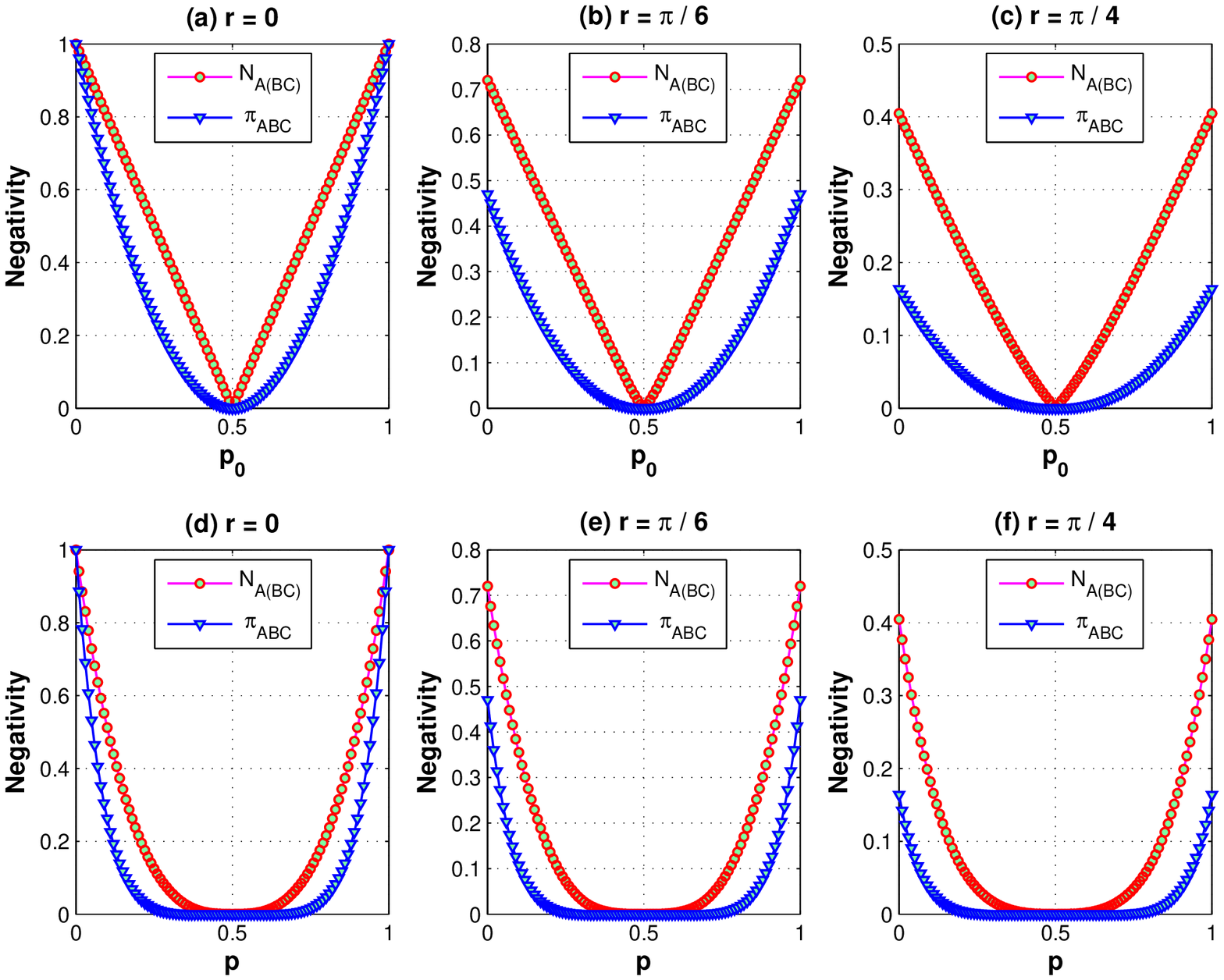} \\[0pt]
\end{center}
\caption{(Color online). The one-tangles and $\protect\pi $-tangles are
plotted as a function of decoherence parameter, $p$ for different values of
acceleration $r$\ for phase flip channel.}
\end{figure}

\begin{figure}[tbp]
\begin{center}
\vspace{-2cm} \includegraphics[scale=0.8]{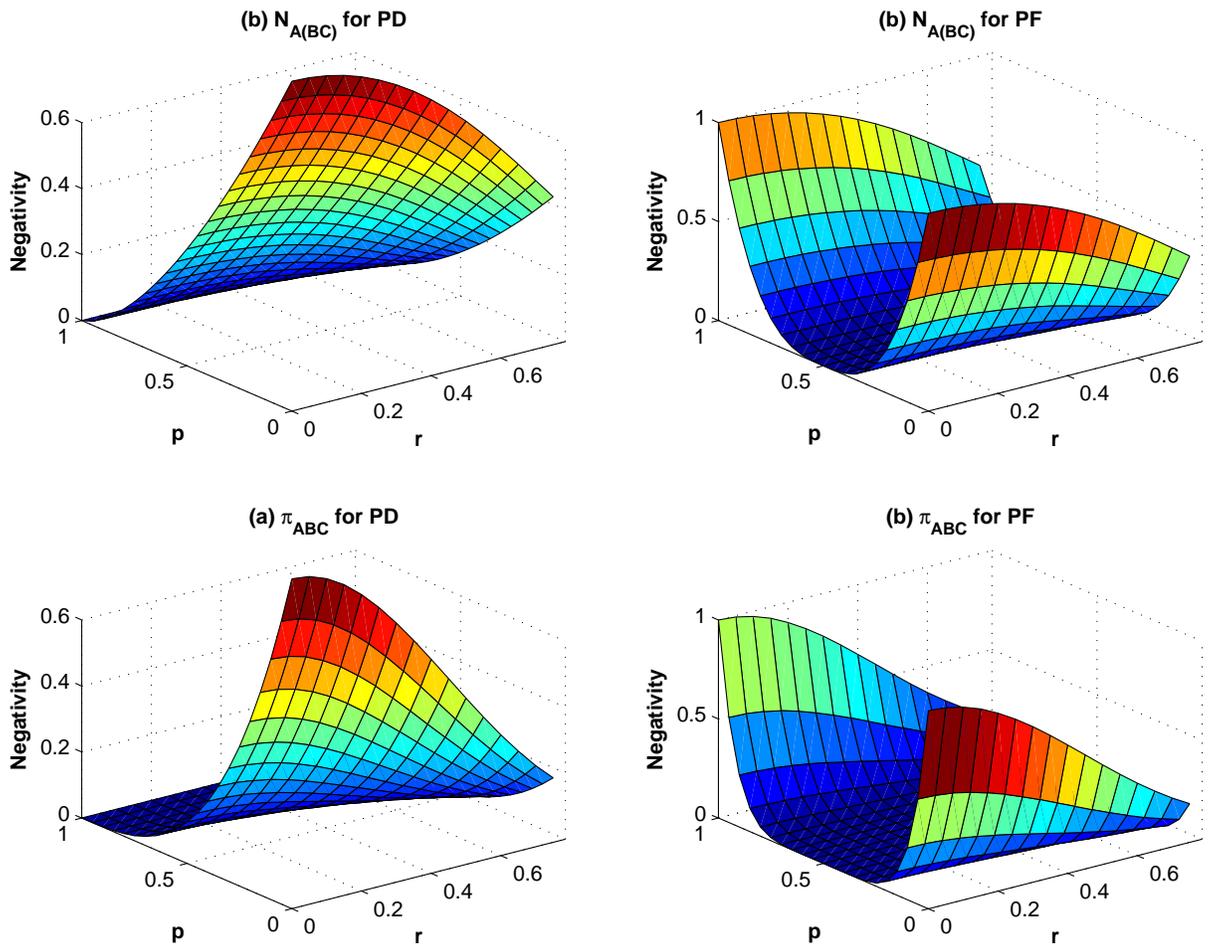} \\[0pt]
\end{center}
\caption{(Color online). The one-tangles and $\protect\pi $-tangles are
plotted as a function of acceleration, $r$\ and decoherence parameter, $p$
for phase damping and phase flip channels.}
\end{figure}

\end{document}